# Sinestesia as a model for HCI: a Systematic Review

Simona Corciulo[1], Mario Alessandro Bochicchio[2]

[1] Department of Computer Science, University of Turin, Italy
[2] Department of Computer Science, University of Bari Aldo Moro, Italy

**Abstract**
Synesthesia, conceived as a neuropsychological condition, may prove valuable in studying the interaction between humans and machines by analyzing the co-occurrence of sensory or cognitive responses triggered by a stimulus. In our approach, synesthesia is elevated beyond a mere perceptual-cognitive anomaly, offering insights into the reciprocal interaction between humans and the digital system, steering novel experimental design and enriching results interpretations. This review broadens the traditional scope, conventionally rooted in neuroscience and psychology, by considering how computer science can approach this condition. The interdisciplinary examination revolves around two primary viewpoints: one associating this condition with specific cognitive, perceptual, and behavioral anomalies, and the other acknowledging it as a prevalent human experience. Synesthesia, in this review, emerges as a significant model for Human Computer Interaction (HCI). The exploration of this specific condition aims to decipher how atypical pathways of perception and cognition can be encoded, empowering machines to actively engage in processing information from both the body and the environment. The authors attempt to amalgamate findings and insights from various disciplines, fostering collaboration between computer science, neuroscience, psychology, and philosophy. The overarching objective is to construct a comprehensive framework that elucidates how synesthesia and anomalies in information processing can be harnessed within HCI, with a particular emphasis on contributing to digital technologies for medical research and enhancing patient's care and comfort. In this sense, the review endeavors also to fill the gap between theoretical understanding and practical application.

## 1. Introduction

From a neuroscientific perspective, sensory information coming from inside our body (interoceptive) and from the outside (exteroceptive) are processed through different mechanisms and can be distinctly perceived. However, both contribute to the construction of our sensory experience and the perception of ourselves and the environment. Understanding the distinction between these two information sources is crucial for comprehending how the brain interprets and integrates sensory signals (Craig, A. D. (2002). How do you feel? Interoception: the sense of the physiological condition of the body. Nature reviews neuroscience, 3(8), 655-666.)

In our daily experience, perceptions shape our sensations that contribute to the definition of our feelings, moods and memories and these, in turn, influence our subsequent perceptions. This continuous flow, which places our consciousness in intimate contact with what we can perceive of the outside world and other minds, shapes our own consciousness, and then brings back into the world our thoughts transformed into deeds, words, and moods. Sciences and Philosophy have devoted considerable attention to the study of these phenomena, providing a solid foundation for Neuroscience and Information Science, which, in recent years, have developed theories, tools and conceptual models that have made important contributions to the study of the relationship between perception and awareness. Artificial Intelligence (AI), Human Computer Interaction (HCI), and many other disciplines and technologies have received decisive impulses from these studies.

Synesthesia is a neuropsychological condition affecting 2%-4% of the population (Simner J, Sagiv N, Mulvenna C, Tsakanikos E, Witherby S. A, et al. Synaesthesia: the prevalence of atypical cross-modal experiences. Perception. 2006;35(8):1024-1033.) in which stimuli presented through one modality spontaneously evoke conscious sensations in an unrelated modality, and this is perceived as an "unconventional sensation" by affected subjects. The simultaneity of sensory and cognitive responses elicited by a given stimulus, combined with the possibility of using digital systems to activate a wide range of stimuli, offers a unique lens for exploring how perceptions are related to conscious sensations. This, in turn, can contribute to a better understanding of the dynamics of the human-machine relationship, can help design more effective and personalized digital applications, and can improve our knowledge of information processing mechanisms within human and artificial systems.

This interdisciplinary examination is structured around two main perspectives: one linking synesthesia with specific cognitive, perceptual, and behavioral anomalies, and another considering it as a widespread human experience, not limited to synesthetes but shared by all subjects. From this perspective, synesthesia is treated as a significant model for Human-Computer Interaction (HCI), possibly enabling new interaction paradigms between machines and humans in various contexts.

The exploration of this condition aims to decipher how atypical sensory and cognitive pathways can be encoded, enabling machines to actively participate in processing information originated from both the body and the environment. The overall goal is to understand how synesthesia and anomalies in information processing can be leveraged in the field of HCI, with a particular focus on developing technologies for medical research and improving patients' care and comfort.

In the context of cognitive sciences and information technology, theories and research continually redefine our understanding of information and consciousness. This review explores three fundamental areas that together provide an integrated narrative on how we perceive and process information and how this relates to the interaction between humans and machines.

Section I introduces Shannon's information and Tononi's integrated information theories, separated by a sixty-year interval, providing essential tools to understand the transfer and integration of information and how these can help our understanding of perception, cognition, and consciousness. The improvement of our (still incomplete) understanding of the mechanisms regulating consciousness is crucial for the medical field, helping to understand what makes people conscious or not when experiencing something, helping the development of systems that not only respond to medical needs but are also sensitive to patients' conscious experiences, thus improving their care and comfort. In this perspective, synesthesia can serve as a bridge between the theory of consciousness and practical applications in HCI, with a focus on technologies supporting medical research and patients' comfort and care.

Section II delves into the concept of error in information transmission, both in the context of traditional information theory and the human perceptual system, examining how perceptual errors influence our awareness and how consciousness is shaped not only by accurate sensory information but also by interpretations and corrections applied by the brain to sensory stimuli. Section II further explores the concept of error in information transmission, both in the context of traditional information theory and the human perceptual system. This analysis is crucial for HCI, as perceptual 'errors' are not simple anomalies to correct but rather opportunities to understand how consciousness is shaped. Synesthesia, in this context, becomes a paradigmatic example: the atypical 'mappings' of stimuli to sensory responses in synesthetic individuals illuminate the pathways through which consciousness interprets the world. This understanding can be harnessed in HCI to create technologies that not only handle errors more humanely but also adapt and respond to the diverse nature of human experience, especially in medical contexts where considerations about the patient's perceptions and sensations are crucial.

Section III focuses on the remarkable adaptive capacity of the human brain, described in terms of plasticity during childhood and adulthood. It examines how this capacity can be supported and optimized through the use of devices designed to model cognitive and sensory functions, documenting the effectiveness of such procedures in various fields, especially in the recovery of higher-order abilities and executive functions compromised by aging. Additionally, it evaluates how studies conducted with sensory-deprived adults challenge traditional assumptions about the factors guiding the sensory organization of the brain, especially in relation to critical developmental periods, and the implications of such studies in the context of rehabilitative approaches for sensory restoration and their impact on residual sensory abilities. This research could have implications for rehabilitative approaches based on HCI, highlighting how the model offered by synesthesia can improve residual sensory abilities, sensory restoration and personalized medical assistance.

## 1.1 Research questions

This exploration is framed by three fundamental questions: (i) How does synesthesia uniquely contribute to our understanding of human-machine interaction? (ii) In what manner can the atypical perceptions seen in synesthesia contribute to innovating the dynamics between humans and machines? (iii) What collaborative prospects exist within the realms of computer science, neuroscience, psychology, and philosophy to effectively incorporate synesthesia into Human-Computer Interaction (HCI)? Through an in-depth analysis, this review endeavors to close the divide between theoretical knowledge and real-world application, by exploring the intricate relationship between synesthesia and HCI.

## 1.2 Intended audience

This review could be of interest to researchers, scholars, and professionals in the fields of computer science, especially those involved in human computer interactions. Additionally, it may be valuable for those engaged in technological research and development, with a particular emphasis on clinical applications related to sensory impairment and cognitive decline. The review proposes an interpretation of synesthetic experiences as a key to extend what the tools and methods of human computer interaction can produce.

## 2. Related Works

### Section I

In this section, two fundamental theories that, sixty years apart, have shaped the way we understand and use information are described in parallel: those that travel outside our body in the form of bits and those generated by the bioelectrical activity of our brain. The two theories in question are Shannon's information theory and Tononi's integrated information theory. These theories pertain to the transfer, processing, and integration of information and how they converge in the formation of a coherent and unified system.

The topics addressed in this section lay the groundwork for subsequent investigations into the complex relationship that binds perception, cognition, consciousness, and the way digital applications can be implemented, introducing the role of information and its transmission. It will set the stage for discussions regarding multisensory integration and subjective awareness, a crucial aspect concerning the management and manipulation of human experiences.

The "hard problem" of consciousness is introduced, debating its inherent dependence on the concept of integrated information. Investigations into consciousness could have a significant impact on the design of technologies that involve, alter, and integrate sensory perception, as well as assessing the extent to which synesthesia can be considered a form of efficient information transmission, well beyond the traditional definition.

Finally, the relationship between one of the theories and connectomics, a discipline that deals with mapping and studying the neuronal connections within the brain, is also discussed, because of the possible link between synesthesia and anomalous connections among neural circuits.

*Information Theory*

In 1948, Claude Shannon provided a robust mathematical foundation for the measurement and transmission of data with the introduction of information theory (hereafter IT), publishing "A Mathematical Theory of Communication" (Shannon, C. E. (1948). A mathematical theory of communication. The Bell system technical journal, 27(3), 379-423.). He demonstrated how to compress data before transmission and achieve nearly error-free communication, where "error" refers to the discrepancy between the transmitted and received messages in a communication system. This can be addressed by developing error correction codes that enhance transmission reliability (e.g., Hamming code). IT introduced innovative concepts for measuring the amount of information stored in a string or transmitted through a communication channel. Shannon's entropy is among these concepts; it provides a measure of the uncertainty associated with a set of data or an event. Higher entropy indicates greater uncertainty for message recipients and can be used to identify data sequences with a low amount of information (low entropy), allowing for a more efficient representation (see data compression, Verdu, S. (1998). Fifty years of Shannon theory. IEEE Transactions on information theory, 44(6), 2057-2078.; Xu, Q. (2007). Measuring information content from observations for data assimilation: Relative entropy versus Shannon entropy difference. Tellus A: Dynamic Meteorology and Oceanography, 59(2), 198-209.).

Shortly after the publication of the article, some scientists began exploring how information theory could be applied to neuroscience. These developments led, for example, to the introduction of the concept of neural information flow (Lungarella, M., & Sporns, O. (2006). Mapping information flow in sensorimotor networks. PLoS computational biology, 2(10), e144.; DeWeese, M., & Bialek, W. (1995). Information flow in sensory neurons. Il Nuovo Cimento D, 17, 733-741.), referring to the study of information flow in the context of neural systems. This involves identifying the pathways through which neural signals move between different regions of the brain, how they are modulated during specific tasks, and how they contribute to the generation of behavioral responses.

Timme et al. (2018) (Timme, N. M., & Lapish, C. (2018). A tutorial for information theory in neuroscience. eneuro, 5(3).) explored the application of information theory in neuroscience, focusing on understanding how neural systems process multivariate data with nonlinear interactions. They addressed logistical issues such as data types, data partitioning, quantitative requirements, and biases. To facilitate the use of such analyses, they provided free MATLAB software applicable to a wide range of data from neuroscience experiments and other fields.

*Information Theory and Connectomics*

The brain's functionality relies on efficient communication among its various parts, a process regulated by the connectome, a complex network of structural connections within the nervous system. Connectomics aims to investigate how the organization of the connectome influences the brain's ability to process information and perform specific functions. Analyses of brain networks have revealed common principles of connectivity, such as the minimization of physical and metabolic costs by promoting local circuits (Bazinet, V., Hansen, J. Y., & Misic, B. (2023). Towards a biologically annotated brain connectome. Nature reviews neuroscience, 24(12), 747-760.). Despite the substantial energy demands, connectomes facilitate communication and global integration of information (van den Heuvel, M. P., & Sporns, O. (2019). A cross-disorder connectome landscape of brain dysconnectivity. Nature reviews neuroscience, 20(7), 435-446.). Information theory plays a prominent role in connectomics, providing an essential conceptual framework for understanding the representation and transfer of information in the neural networks of the brain. The primary objective of connectomics is the mapping and comprehension of connections between neurons and brain regions, and Shannon's theory enriches this perspective by introducing key concepts.

Quantifying information is facilitated by IT metrics and concepts, enabling the assessment of the amount of information transmitted through neural connections and contributing to discerning the efficiency and importance of these connections. Entropy analysis is employed to explore the diversity and complexity of information in neural networks, offering insight into the variety of processed information (Wibral, M., Vicente, R., & Lindner, M. (2014). Transfer entropy in neuroscience. Directed information measures in neuroscience, 3-36.; Keshmiri, S. (2020). Entropy and the brain: An overview. Entropy, 22(9), 917.). Information theory allows considerations on the efficiency of neural connections in information transfer, identifying more efficient connections that may indicate neural circuits specialized for specific cognitive functions. IT principles are also incorporated into computational models, enabling the simulation and understanding of how information propagates through specific neural networks and predicting network behavior based on mapped connections (Clark, J. W., Rafelski, J., & Winston, J. V. (1985). Brain without mind: computer simulation of neural networks with modifiable neuronal interactions. Physics Reports, 123(4), 215-273.; Schmidt, R., de Reus, M. A., Scholtens, L. H., van den Berg, L.
H., & van den Heuvel, M. P. (2016). Simulating disease propagation across white matter connectome reveals anatomical substrate for neuropathology staging in amyotrophic lateral sclerosis. Neuroimage, 124, 762-769.). Finally, similar to telematic networks, central nodes or "hubs" in neural networks play a crucial role in the transmission and integration of information. Hubs emerge as key elements for the collaboration of different brain regions in cognitive functions (Van Den Heuvel, M. P., & Sporns, O. (2011). Rich-club organization of the human connectome.

Journal of Neuroscience, 31(44), 15775-15786.; Van den Heuvel, M. P., & Sporns, O. (2013). Network hubs in the human brain. Trends in cognitive sciences, 17(12), 683-696.). Therefore, information theory is one of the indispensable conceptual tools in interpreting connectomics data, providing an in-depth understanding of how information is represented and processed in the brain through neural connections.

Mišić et al. (2016) (Mišić, B., Betzel, R. F., Nematzadeh, A., Goni, J., Griffa, A., Hagmann, P., ... & Sporns, O. (2015). Cooperative and Competitive Spreading Dynamics on the Human Connectome. Neuron, 86(6), 1518-1529.) published a study titled "Cooperative and Competitive Spreading Dynamics on the Human Connectome". The authors used brain imaging data to map the human connectome and applied IT concepts to understand how brain activity spreads through the neural network. Specifically, they explored the dynamics of the spread of brain activity in terms of cooperation and competition between brain regions. They used measures such as entropy and mutual information to quantify cooperation and competition between different brain regions during the information diffusion process. Their study contributed to a deeper understanding of how information propagates through the neural network, identifying patterns of cooperation and competition between brain regions. The combined use of connectomics and information theory has allowed for more detailed insights into the complex dynamics of neural interactions in the human brain.

*Integrated Information Theory (IIT)*

A high degree of information integration may be associated with the process to transform a perceptual stimulus into a unified, sensorial conscious experience. Integrated Information Theory (henceforth IIT) by Tononi, developed in 2004 and revisited in subsequent years, presents an approach to understanding consciousness (and conscious perceptions) based on the concept of integrated information. In IIT, Φ (phi) represents the quantitative measure of integrated information within a system. This measure is central and used to assess the degree of consciousness associated with a system. In simple terms, Φ measures how information is integrated and correlated within a system, indicating the system's capacity to generate a unified and coherent conscious experience: a higher Φ value is associated with greater complexity and cohesion in information, suggesting a higher level of consciousness.

It should be emphasized that Φ does not merely represent a measure of the quantity of information present in the system but is also indicative of the quality of the integration of such information. A system with a large amount of information that is not strongly integrated might have a lower Φ value compared to a system with a more limited but highly integrated amount of information. In this way, Tononi suggests that the human brain, with its complex network of neural connections, is capable of generating high levels of integrated information, giving rise to our rich conscious experience.

Mediano et al. (2022) (Mediano, P. A., Rosas, F. E., Bor, D., Seth, A. K., & Barrett, A. B. (2022). The strength of weak integrated information theory. Trends in Cognitive Sciences.) address IIT, highlighting its extraordinary ambition in proposing a universal mathematical formula, derived from the fundamental properties of conscious experience, to quantify the quality and quantity of consciousness in any physical system. IIT was conceived as a potential solution to the hard problem of consciousness, i.e., understanding why and how physical processes are accompanied by subjective experiences. However, the authors' analysis highlights that, in its current formulation, empirical evidence does not support the level of specificity present in the theory. They explore the divisive nature of IIT, with some seeing it as a powerful approach to the hard problem and others rejecting it as unverifiable. For this reason, they propose the distinction between strong IIT, in which every system showing a significant amount of integrated information manifests some degree of consciousness, even if very basic, assuming that it can be present in variable degrees even in systems other than the human brain, and weak IIT, in which only complex systems, such as the human brain, exhibit a high degree of consciousness, assuming that it is closely tied to advanced biological systems and is not attributed to all systems showing some level of information integration. Furthermore, the authors discuss the challenges associated with strong IIT, explaining how current empirical findings are consistently explained by weak IIT without necessarily fully adhering to strong IIT.

Despite the debate and criticisms, IIT remains one of the most intriguing proposals for understanding the nature of consciousness. The association of information has been identified as a distinctive attribute of human consciousness, as it requires the global availability of information through extensive neural interactions. However, the intricate interconnections between multisensory integration and perceptual awareness, namely consciousness, remain to be defined.

**Section II**
The purpose of this section is to address the concept of synesthesia, intended as a noisy communication between the sense organs and the brain, using the concepts of Information Theory (IT) and Integrated Information Theory (IIT). In this context, "information reintegration" and "diversity gain" approaches, potentially suitable for HCI implementations, have been defined in the literature to cope with the perceptual anomalies of synaesthetes, but also to extend the conscious perception experiences in non-synesthetes.
In information theory, the probability of error refers to the likelihood of an error occurring during the transmission of information through a communication channel. This concept is fundamental when considering the transmission of signals through an environment that can introduce noise, interference or distortions into the signal.
When considering the human neural system, the concept of error is more complex. In the human perceptual system, sensory information from sense organs is processed by the brain to form a perceptual experience. This process can face many interferences (cognitive, environmental, etc.) (Kastanakis, M. N., & Voyer, B. G. (2014). The effect of culture on perception and cognition: A conceptual framework. Journal of Business Research, 67(4), 425-433.; Russell, L. E., Yang, Z., Tan, P. L., Fişek, M., Packer, A. M., Dalgleish, H. W., ... & Häusser, M. (2019). The influence of the visual cortex on perception is modulated by behavioral state. biorxiv, 706010.). In fact, the human perceptual system is involved in a complex network of processes that go beyond simple linear transmission.
Errors in human perception can stem from sensory limitations, cognitive distortions, subjective interpretations, and other causes. Unlike information theory, where error is merely a deviation from the accurate transmission of a signal, in human perceptual experience, error in transmission is a more nuanced phenomenon and intrinsically linked to the complexity of the processing mechanisms involved.
The influence of various error sources in the human perceptual system can have a
significant impact on consciousness. Since consciousness is closely tied to the processing of information from sense organs and cognitive interpretations, perception errors can shape our awareness of the surrounding world. In the presence of sensory errors or cognitive distortions, the brain might generate distorted or erroneous perceptions of reality. These errors can contribute to optical illusions, auditory misunderstandings, or incorrect perceptions of events. Goodie and Fortune (2013) explored cognitive distortions in pathological gambling, highlighting the need for precise measurement tools to better understand this phenomenon. Zavagno, Daneyko, and Actis-Grosso (2015) discussed the conceptualization of perceptual illusions, emphasizing how errors and cognitive mishaps contribute to their formation, providing a unique perspective on the cognitive processes underlying incorrect visual experiences. Both publications offer significant insights into cognitive mechanisms affecting behaviors and perceptions.(Goodie, A. S., & Fortune, E. E. (2013). Measuring cognitive distortions in pathological gambling: review and meta-analyses. Psychology of Addictive Behaviors, 27(3), 730.; Zavagno, D., Daneyko, O., & Actis-Grosso, R. (2015). Mishaps, errors, and cognitive experiences: On the conceptualization of perceptual illusions. Frontiers in Human Neuroscience, 9, 190.). Consciousness, being the product of complex neural processes, reflects the information it receives and can adapt or be influenced by perceptual errors. The impact on consciousness can range from minor perception errors to more complex situations where significant errors might lead to misunderstandings of reality (Gilchrist, A. (2003). The importance of errors in perception. Colour perception: Mind and the physical world, 435-451.; Firestone, C., & Scholl, B. J. (2016). Cognition does not affect perception: Evaluating the evidence for "top-down" effects. Behavioral and brain sciences, 39, e229.). This highlights the dynamism and subjectivity of consciousness, shaped not only by accurate sensory information but also by interpretations and corrections that the brain applies to sensory stimuli.

In the human perceptual system, the Integrated Information Theory (IIT) approach would suggest that consciousness is not only tied to the accuracy of sensory information but also to the complexity of neural connections contributing to form a unified and integrated experience. Errors or distortions in sensory information can be considered in terms of reduced integrated information and, consequently, reduced consciousness related to those particular pieces of information. Anomalous perceptions can be addressed within information theory by considering how the perceptual system handles unusual or atypical information. These phenomena can be interpreted as deviations from the normal transmission and processing of sensory information. Then, analyzing anomalous perception phenomena could contribute to a deeper understanding of the complexity of information representation in the human brain. These cases offer opportunities to explore how the perceptual system adapts or interprets unconventional information, paving the way for research exploring the flexibility and adaptability of the brain in the context of unusual perceptions (Kolb, B., Gibb, R., & Robinson, T. E. (2003). Brain plasticity and behavior. Current directions in psychological science, 12(1), 1-5.; Kolb, B., & Gibb, R. (2011). Brain plasticity and behavior in the developing brain. Journal of the Canadian Academy of Child and Adolescent Psychiatry, 20(4), 265.; Kujala, T., & Näätänen, R. (2010). The adaptive brain: a neurophysiological perspective. Progress in neurobiology, 91(1), 55-67.). Exploring anomalous perception phenomena could also provide fertile ground for understanding and addressing potential reintegrations or adaptations in the context of HCI. The unusual nature of these perceptions could offer valuable insights into the plasticity of the perceptual system and its ability to handle stimuli. Working on reintegration could involve strategies to help the brain coherently interpret or adapt information from anomalous perceptions, fostering a better understanding and integration of perceptual experiences. This approach could be relevant in clinical, educational, or research contexts, where understanding how the brain handles stimuli can have significant implications.

To analyze and, potentially, solve the problem of anomalous perceptions using HCI techniques, an alternative approach is suggested by Shannon's theory, which, in wireless communications, exploits the concept of spatial diversity to solve the interference problem. The goal is to leverage the diversity of signal propagation conditions in space, reducing the negative effects of interferences, attenuations, and other disturbances.

Appliyng diversity to human beings' perceptions, when information cannot be transmitted on a specific sensory channel, another comes into play (Tarokh, V., Seshadri, N., & Calderbank, A. R. (1998). Space-time codes for high data rate wireless communication: Performance criterion and code construction. IEEE transactions on information theory, 44(2), 744-765.). This approach, called "Diversity gain", refers to the benefit gained in improving information transmission by choosing multiple sensory channels or diversified paths. When a communication channel is subject to disturbances, interferences, or other unfavorable conditions that might compromise reliable information transmission, system robustness can be enhanced by introducing diversity. In IT, this diversity can take various forms, such as using multiple frequencies, more antennas, or multiple transmission paths. The fundamental concept is that, in the presence of disruptions on a particular channel, other channels or paths may be less affected by the same disturbances. Therefore, simultaneous transmission over multiple channels or through different paths can reduce the risk of errors and improve the overall reliability of communication. Similarly, in diversity gain, when sensory stimuli traverse damaged neural circuits and are subject to interference, the appropriate selection of diverse sensorial channels or paths could increase the likelihood of accurate information transmission.

**Fusion of the Senses**

Traditionally, synesthesia is described in terms of the 'union of the senses' or 'fusion of the senses':

- "Synesthesia is a **union of the senses**" (R. Cytowic, Synesthesia, 2002.).
- "Vasilly Kandinsky (1866-1944) had perhaps the deepest sympathy for **sensory fusion**, both synesthetic and as an artistic idea. He explored harmonious relationship between sound and color and used musical terms to describe his paintings , calling them "compositions" and "improvisations." His own 1912 opera, Der Gelbe Klang ("The Yellow Sound"), specified a compound mixture of color, light, dance, and sound typical of the

> Gesamtkunstwerk" (Cytowic, R. E. (1995). Synesthesia: Phenomenology and neuropsychology. Psyche, 2(10), 2-10.).

Synesthesia is considered a rare neurological trait that induces unusual experiences, often involving interconnected senses. There are over a hundred forms of synesthesia (Ward, J., & Simner, J. (2020). Synesthesia: The current state of the field. In Multisensory perception (pp. 283-300). Academic Press.), many of which concern perception. An example of perceptual synesthesia could be the perception of colors associated with sounds. For instance, a synesthetic person might experience, at a conscious level, the vision of a specific color when hearing a particular musical note or sound. This example illustrates why synesthesia is often described as a "fusion of the senses": one sense (the inducer, in this case, sound) merges with another (the concurrent, in this case, vision). This means that the sensory experience extends beyond the sense of hearing, involving vision in a way that goes beyond normal perception. The variety of synesthetic phenomena suggests that the definition of synesthesia goes beyond the traditional one. Stimuli triggering synesthetic experiences can be more abstract, such as thoughts or reasoning. Additionally, synesthetic experiences may involve highly conceptual stimuli, such as personality traits. Sequence-personality synesthesia, for example, associates ordered sequences, like the letters of the alphabet, with specific personality traits (Sobczak-Edmans, M. (2013). A systematic study of personification in synesthesia: Behavioral and neuroimaging studies (Doctoral dissertation, School of Social Sciences Theses); Simner, J., Gärtner, O., & Taylor, M. D. (2011). Cross-modal personality attributions in synesthetes and non-synesthetes. Journal of Neuropsychology, 5(2), 283-301.). These abstract aspects of thinking challenge the idea of synesthesia as a simple "fusion of the senses".

The diversity of synesthetic conditions poses a challenge for diagnostic methods. While grapheme-color synesthesia has known neural correlates, other types of synesthesia may remain less understood. Magnetic resonance imaging, though useful for some types, is not widely used for diagnosis, mainly due to its complexity and associated costs. Although synesthesia is likely inherited, there is no definitive genetic test, as the genetic understanding of this phenomenon is still in its infancy. Consequently, the "gold standard" for diagnosing synesthesia is currently a behavioral test. This test relies on the stability of associations over time, as an individual's synesthetic experiences tend to be consistent over time. However, it has been shown that this is not always the case. Research on synesthesia also extends to investigating its neurological origins and the early development of these particular sensory associations. Understanding how these connections develop in the brain and how they influence daily experience remains a fascinating field of inquiry. Jewanski et al. (2020)(Jewanski, J., Simner, J., Day, S. A., Rothen, N., & Ward, J. (2020). The "golden age" of synesthesia inquiry in the late nineteenth century (1876-1895). Journal of the History of the Neurosciences, 29(2), 175-202.) discuss whether synesthesia should be considered a pathology or an alternative manifestation of intelligence, also examining the roles of heredity and environment. Interpretations of the neural basis of synesthesia have been diverse, raising the question of whether synesthetic experiences are something special or a more intense manifestation of a general ability to form associations.

**Section III**

To be added soon.

**3. Methods**

**3.1 The PRISMA 2020 protocol**

This systematic review on synesthesia and human-machine interaction is based on the PRISMA 2020 protocol (Page, M. J., McKenzie, J. E., Bossuyt, P. M., Boutron, I., Hoffmann, T. C., Mulrow, C. D., ... & Moher, D. (2021). The PRISMA 2020 statement: an updated guideline for reporting systematic reviews. Bmj, 372.) by defining the eligibility criteria, curating the research strategy, systematically organizing and synthesizing the collected data, and subsequently presenting a comprehensive summary of the main findings.

**3.2 The Eligibility Criteria**

The two eligibility criteria agreed upon by the authors are defined and motivated in the following subsections.

**3.2.1 First Criterion**

The first criterion has been elicited by the authors' assumptions concerning the recognition of two main research approaches to the synesthetic condition and primarily regarding the boundaries of its definition. The authors report that one type of definition is principally treated as solid and sufficiently informative, while the other is comparatively and apparently less restrictive and emergent among the researchers. The first is commonly detectable in studies that explore the phenomenon in connection with some cognitive or perceptual deviations. On the other hand, the second defines the synesthetic perception as a condition that can be variously experienced by everyone, rejecting any strict limitations of the phenomenon within the edges of some cognitive, perceptual, or behavioral anomalies.

To achieve a clearer comprehension of these two main definitions, the authors consider other crucial aspects that arise as supportive to the investigation, elements that concern the phenomenon's convolutions and that can be discussed to hack and deeply grasp the intricacies of that condition. Notions and considerations inherent to individuals' perceptual and cognitive abilities and about the dilemma concerning how individuals acquire consciousness about their experiences, are fundamental to approach synesthesia and to resize its shape in view of a computational perspective. The three key dimensions are deemed crucial in the broad domain of human-machine interaction, since considerations related to making individuals aware or unaware of what they see, hear, touch, smell, taste, and emotionally experience are significantly of interest. Thus, the authors found it beneficial to include studies on synesthesia and how these have been approached across diverse disciplines, with a preference for studies produced in neuroscience, computer science, psychology and philosophy.

**3.2.2 Second Criterion**

The second criterion is designed to explore synesthesia within the realm of human-machine interactions. Specifically, the authors aim to introduce a core perspective that regards synesthesia as a mechanism enabling technology to actively participate in perceiving and processing information derived from both the environment and the individual's body. The authors chose to consider studies that assess the significance of multimodal interactions between humans and machines as underlying human perception since they are connected to accessing data and sensing the world by combining different modalities. They decided to include studies that either support or challenge current knowledge in neuroscience and psychology, considering them fundamental to embracing new methodological complexities arising from the type of interaction highlighted in the profiled review. The peculiar phenomenon and its inherent neuropsychological functioning are considered by the authors as profoundly linked to some pure computational dimensions: communication, information transmission, and data structuring are described as elements bridging synesthesia to debates about how people can improve or recover their abilities to grasp the realm of environmental data and how this can be impacted by digital systems. This entails perceiving information as communicable through a channel, susceptible to potential interferences, interruptions, alterations, or substitutions in artificial systems collecting and processing information. Analogously, this occurs in the human perceptual system and its associated phases of processing perceived data and stimuli. The strategy of dual inclusion was devised to offer a nuanced understanding of the complex interaction between various types of perceptions within the broader definition of synesthesia, inspiring potential applications in computer research and development. The research toward medical applications was guided by inquiries into the practicality of the expansive definition of synesthesia and its already-recognized efficacy in treating affected patients. However, they evaluated this emphasis as particularly discernible in contexts involving the observation of patients with disabilities. The incorporation of pertinent observations is extended to comparative analyses between synesthetes and non-synesthetes, as well as between individuals with and without disabilities. This comprehensive approach was structured to capture the nuances of synesthetic experiences in specific human-machine interaction scenarios. The pronounced emphasis on technological advancements was noted, especially in clinical contexts where the integration of alternative stimuli or experiential pathways aimed at improving patient well-being plays a central role. By adopting this multifaceted and detailed approach, the goal was to enrich the understanding of synesthetic experiences and their implications within the evolving landscape of human-computer interaction, with particular attention to the efficient management of data, information, and interactions, especially within the realm of clinical applications. Thus, the authors found it beneficial to include studies on synesthesia and how these have been approached across diverse disciplines, with a preference for studies produced in computer science and the domain of medical applications.

### 3.2 The research strategy

The authors focus exclusively on English-language articles available in full over the past two decades. The select sources encompass the Association for Computing Machinery (ACM), Scopus, PubMed, the Neuroscience Information Framework (NIF), the American Psychological Association (APA) and Google Scholar. This diverse array of sources allowed for a thorough examination of synesthesia literature spanning various academic disciplines from 2004 to 2024. In their pursuit of understanding and exploring the multifaceted phenomenon of synesthesia, the researchers have formulated a series of targeted queries to navigate the vast landscape of scholarly literature. These queries are strategically designed to unravel the intersections between synesthesia and diverse academic domains, shedding light on its implications in neuroscience, psychology, computer science, and philosophy. Each query is meticulously crafted with logical operators, such as 'AND' and 'OR,' to refine the exploration and extract nuanced insights from academic studies:

- "Synesthesia" OR "Synaesthesia"     AND "Neuroscience" OR "Psychology" OR "Computer Science" OR "Human-Computer Interaction" OR "Philosophy";
- "Synesthesia" AND "Perception" OR "Cognition" OR "Consciousness";
- "Synesthesia" AND "Human-Computer Interaction" AND "Consciousness" OR "Perception" OR "Cognition";
- "Synesthesia" AND "Synesthete" OR "Not-Synesthete";
- "Synesthesia" AND "Multimodality" AND "Computer Science" AND "Data";
- "Synesthesia" AND "Virtual Reality" OR "VR" OR "Disability" OR "Patient Care".

The presented queries aim to explore various aspects of synesthesia across multiple disciplines. The first set of queries focuses on investigating the relationship between synesthesia and many academic fields. The second set delves into how synesthesia intersects with perception, cognition, and consciousness, highlighting its role in shaping human experiences. Another group of queries examines the connection between synesthesia and human-computer interaction, particularly concerning consciousness, perception, and cognition. Additionally, queries are designed to distinguish between synesthetes and non-synesthetes, shedding light on differences in sensory experience. Furthermore, queries explore the integration of synesthesia with multimodality and computer science, emphasizing its relevance in data-related contexts. Lastly, queries address the implications of synesthesia in virtual reality environments, including its potential impact on individuals with disabilities and its role in enhancing patient care. They applied the following filters: English language, publication year from 2010 to 2020, document type article. The search yielded * records, of which * were removed due to duplication. We applied the following inclusion and exclusion criteria to select relevant studies:

(a) inclusion: studies reporting empirical data on synesthesia, involving both synesthetic and non-synesthetic subjects, using quantitative or qualitative methods;
(b) exclusion: studies not reporting empirical data but only reviews, opinions, comments, or clinical cases; studies not focusing on synesthesia but on other sensory or cognitive phenomena; studies not accessible in full-text format.

We excluded * records based on the exclusion criteria, providing the reason for exclusion for each record. We included * studies in the systematic review, of which * were also included in the meta-analysis.

**Search strategy**

We have provided a detailed description of the search strategy employed to identify relevant studies on synesthesia within the three selected databases: Google Scholar, PubMed, and ResearchGate.

**Selection process**

We have provided a detailed description of the process for selecting relevant studies on synesthesia, applying pre-defined inclusion and exclusion criteria. The EndNote software was used to remove duplicate records and manage the records throughout the selection process. Two reviewers independently examined the titles and abstracts of the records obtained from the search, selecting those potentially relevant for the review. The full texts of the selected records were then retrieved and evaluated by two reviewers according to the inclusion and exclusion criteria. Any discrepancies between the reviewers were resolved through discussion or consultation with a third reviewer. We recorded the number of excluded records and the reason for exclusion at each stage of the selection process. The number of studies included in the systematic review was reported, along with a list of included studies and their characteristics. The selection process is depicted in the PRISMA 2020 flowchart (Figure 1).

**Quality assessment**

We assessed the methodological quality of the studies included in the systematic review, employing the STROBE checklist evaluates the completeness of reporting with 22 items related to the title, abstract, introduction, methods, results, and discussion of observational studies. Two reviewers independently applied the quality assessment tools to the included papers and resolved discrepancies through discussion or consultation with a third reviewer. We reported the number of satisfied STROBE items for each paper in Table 4.

**Data synthesis**

We synthesized the data from the papers included in the systematic review using both narrative and quantitative methods. For the narrative synthesis, we described the characteristics, methods, and results of the papers, Tables and graphs were utilized to summarize key information and facilitate comparison between papers.

**Future Research**

In the landscape of future research to enhance patient care and comfort, technological innovation is proving to be a valuable ally. Emerging applications, such as virtual reality for chronic pain treatment, telemedicine for remote health condition management, and stress management apps, are revolutionizing healthcare. Specifically, synesthesia, with its ability to merge perceptions, offers an intriguing model for new applications that enhance patient comfort. These may include the creation of immersive multisensory environments, the development of intuitive user interfaces that respond in synesthetic ways and the transformation of clinical data into visual or auditory representations to facilitate patient disease comprehension. A specific example is the conversion of sound emissions from MRI scanners into visual representations, which could significantly reduce anxiety during scans, improving the overall experience. Future research should continue to explore these possibilities, developing solutions that are technically feasible and enrich the human experience by leveraging our ability to perceive the world in rich and varied ways. This interdisciplinary approach could lead to significant improvements in how patients experience healthcare, making procedures less intimidating and more welcoming.

**Conclusion**

In conclusion, synesthesia, conceived as a neuropsychological condition, emerges as a significant opportunity to better understand human-computer interaction (HCI), providing new interaction paradigms and a deeper understanding of atypical perceptions. This interdisciplinary review has highlighted how anomalies in perception and information processing can influence technological development and enhance patient care and comfort in some medical fields. This work lays the foundation for future research on the intricate interplay between synesthesia and HCI, with meaningful implications for professionals and researchers engaged in human-machine interaction and clinically-oriented technological development.

In reality, design within the fields of Human-Computer Interaction (HCI) and User Experience (UX) Design is heavily user-focused. User-Centered Design (UCD) is a foundational principle in these areas, emphasizing the importance of understanding users' needs, preferences, and behaviors in the design process. This approach encourages designers to conduct user research, usability testing, and design iterations based on user feedback to create products that truly meet their needs and enhance their experience. While traditional views might assume some uniformity in sensory perception, modern HCI and UX designers recognize and value the diversity of users, including different learning styles, abilities, and limitations. Furthermore, accessibility and inclusivity have become increasing priorities, leading to the design of solutions that accommodate a wide range of sensory experiences and cognitive abilities. Therefore, although the analysis of synesthesia may highlight the variety of perceptual experiences and the need to consider these differences, it's important to acknowledge that modern practices in HCI and UX are already deeply committed to customizing and adapting interfaces to meet the individual needs of users.

In summary: beyond its nature of neuro-sensory anomaly, synesthesia has emerged as the polysemous sensory tool that not only synesthetes, but all of us, use to express our inner meanings and produce effective multimodal communication.